\documentclass[twocolumn]{jpsj2} %% two-column layout
\usepackage{bm}

\title{Antiferromagnetic Nuclear Resonance in the Quasi-Two-Dimensional (CuBr)LaNb$_2$O$_7$ }

\author{Makoto \textsc{Yoshida}\thanks{E-mail address: yopida@issp.u-tokyo.ac.jp}, 
Nobuyuki \textsc{Ogata}, 
Masashi \textsc{Takigawa}\thanks{E-mail address: masashi@issp.u-tokyo.ac.jp}, 
Taro \textsc{Kitano}$^{1}$, 
Hiroshi \textsc{Kageyama}$^{1}$, 
Yoshitami \textsc{Ajiro}$^{1}$, 
and Kazuyoshi \textsc{Yoshimura}$^{1}$}

\inst{Institute for Solid State Physics, University of Tokyo, Kashiwa, Chiba 277-8581 \\ 
$^{1}$Department of Chemistry, Graduate School of Science, Kyoto University, Kyoto 606-8502}

\abst{We report nuclear magnetic resonance (NMR) studies in the antiferromagnetic state of 
the quasi-two-dimensional (CuBr)LaNb$_2$O$_7$.  The NMR spectra at zero magnetic field and 
4.2~K indicate a unique Cu and Br sites with an internal field of 5.7~T (at Cu) and 16.4~T (at Br), 
confirming a magnetic order.  For the large internal field at the Br sites to be compatible with 
the collinear antiferromagnetic order observed by neutron diffraction experiments 
(N. Oba {\it et al.}, J. Phys. Soc. Jpn. {\bf 75}, (2006) 113601), the Br atoms must move 
significantly off the center of the square of the Cu sublattice so that the Br nuclei couple
predominantly to two parallel Cu moments.  While invalidating the frustrated 
$J_{1}$-$J_{2}$ model defined on a $C_{4}$-symmetric square lattice, our results are  
compatible with the structural model proposed for (CuCl)LaNb$_2$O$_7$ by Yoshida {\it et al.} 
(J. Phys. Soc. Jpn. {\bf 76}, (2007) 104703).}

\kword{NMR, (CuBr)LaNb$_2$O$_7$, quantum spin system, collinear order, structural distortion}

\begin{document}
\maketitle

\section{Introduction} %% No sections necessary for express letters, letters and short notes 
The recent successful synthesis of a new Dion-Jacobson series of quasi-two-dimensional 
quantum spin systems (Cu$X$)$A_{n-1}B_{n}$O$_{3n+1}$ ($A$ = La$^{3+}$, Ca$^{2+}$, 
Na$^{+}$, ..., $B$ = Nb$^{5+}$, Ta$^{5+}$, Ti$^{4+}$, $X$ = Cl, Br, $n$ = 2, 3, 4, ...) by ion-exchange 
reaction has enabled us to explore a wide variety of novel quantum phenomena in a common 
structure of the Cu$X$ magnetic layers, where Cu$^{2+}$ ions (spin 1/2) form a square lattice 
with $X^{-}$ ions located at the center of the square~\cite{Kodenkandath991,Kageyama051,
Kageyama052,Oba061, Tsujimoto071,Kitada071,Yoshida071}.  For example, (CuCl)LaNb$_2$O$_7$ 
has a singlet ground state with an excitation gap of 2.3~meV~\cite{Kageyama051} 
and shows an intriguing phase transition in magnetic fields~\cite{Kageyama052,Kitada071,Yoshida071}. 
On the other hand, a collinear antiferromagnetic order at the wave vector $\boldsymbol{Q}
= (\pi, 0, \pi)$ occurs below $T_{N}$ = 32~K in (CuBr)LaNb$_2$O$_7$~\cite{Oba061}. 
Furthermore, (CuBr)Sr$_2$Nb$_3$O$_{10}$ with a wider interlayer separation exhibits a 
puzzling magnetization plateau at 1/3 of the saturation~\cite{Tsujimoto071}. 

\begin{figure}[b]
\begin{center}
\includegraphics[width=0.9\linewidth]{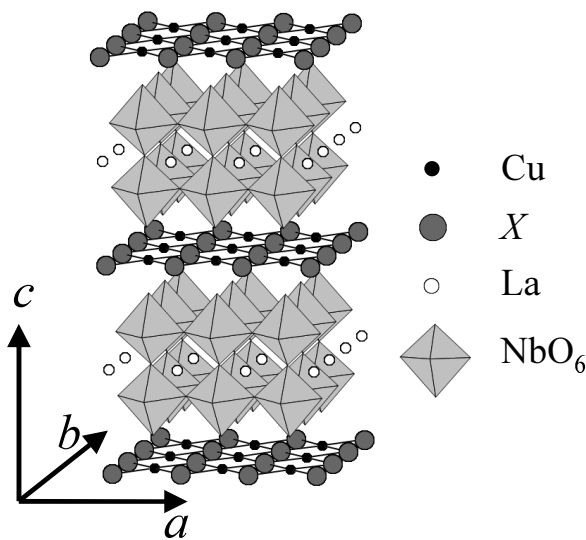}
\end{center}
\caption{Schematic crystal structure of (Cu$X$)LaNb$_2$O$_7$ ($X$ = Cl, Br)~\cite{Kodenkandath991}.}
\label{structure}
\end{figure} 
Early structural studies on (Cu$X$)LaNb$_2$O$_7$ ($X$ = Cl or Br) reported 
the tetragonal $P4/mmm$ space group, where both the Cu and the $X$ sites have the $C_{4}$ 
symmetry~\cite{Kodenkandath991,Kodenkandath011} (Fig.~1).  Based on this, a Heisenberg 
spin model with the nearest neighbor exchange $J_{1}$ and the second nearest neighbor 
exchange $J_{2}$ (the so-called $J_{1}$-$J_{2}$ model) was considered an appropriate 
model. In fact, extensive theoretical work on the 
frustrated $J_{1}$-$J_{2}$ model predicts a collinear antiferromagnetic order with the wave 
vector $(\pi, 0)$ for $J_{2} \gg |J_{1}|$ and a spin-singlet phase in the highly frustrated region 
$J_{2}/ |J_{1}| \sim 0.5$ both for the positive (antiferromagnetic)~\cite{Misguich031,Read891} 
and the negative (ferromagnetic)~\cite{Shannon041} values of $J_{1}$.  This appears to 
be corroborated nicely by the experimental observation of the collinear order in 
(CuBr)LaNb$_2$O$_7$ and the singlet ground state in (CuCl)LaNb$_2$O$_7$. The analysis 
of the susceptibility and the magnetization process indicated that $J_{1}$ is ferromagnetic in 
(Cu$X$)LaNb$_2$O$_7$.

\begin{figure}[t]
\begin{center}
\includegraphics[width=1\linewidth]{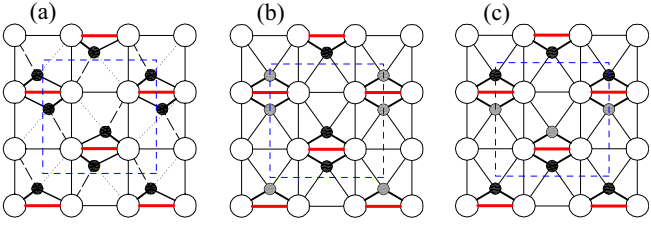}
\end{center}
\caption{(color online).  Possible structures of the CuCl plane proposed for 
(CuCl)LaNb$_2$O$_7$~\cite{Yoshida071}.  The open and solid circles 
represent the Cu and Cl atoms, respectively. In (a), The Cl atoms 
moves from the ideal (0, 0, 1/2) position to ($x$, $y$, 1/2), ($-x$, $y$, 1/2), ($x$, $-y$, 1/2), 
or ($-x$, $-y$, 1/2) and are still on the same plane as the Cu atoms. In (b) and (c), a half of the 
Cl atoms moves up to ($x$, 0, $1/2+\delta$) or ($-x$, 0, $1/2+\delta$) (black circles) 
and another half moves down to ($x$, 0, $1/2-\delta$) or ($-x$, 0, $1/2-\delta$) (grey circles). 
The enlarged unit cell is shown by the blue dashed lines. Note that both Cu and Cl occupy a unique 
crystallographic sites for all cases.  The singlet dimer bonds are shown by the thick red lines.}
\label{model}
\end{figure} 
However, structural information obtained from subsequent experiments 
has raised serious doubt against the validity of the $J_{1}$-$J_{2}$ model. The 
neutron diffraction experiments on (CuCl)LaNb$_2$O$_7$~\cite{Caruntu021} suggested 
significant deviation of the Cl position from the center of the Cu square along the 
[100] direction. The Synchrotron X-ray diffraction on (FeCl)LaNb$_2$O$_7$~\cite{Oba071} 
and the transmission electron microscopy (TEM) measurements on 
(CuCl)LaNb$_2$O$_7$~\cite{Yoshida071} revealed a superlattice reflections indicating 
an enlarged $2a \times 2b \times c$ unit cell.  The nuclear magnetic 
resonance (NMR) and the nuclear quadrupole resonance (NQR) experiments in 
(CuCl)LaNb$_2$O$_7$~\cite{Yoshida071} demonstrated absence of the four-fold symmetry 
around the $c$-axis at both the Cu and the Cl sites.  In particular, the principal axis 
of the electric field gradient (EFG) tensor at the Cu sites with the largest principal value 
was found to be perpendicular to the $c$-axis, providing conclusive evidence for a drastic 
change of the local structure from the tetragonal $P4/mmm$ symmetry.  Yet observation of a 
single sharp NQR line for the Cu and the Cl nuclei guarantees that they both occupy a unique 
crystallographic site without significant disorder.  Yoshida \textit{et al.} has proposed 
possible structural models of the CuCl planes in (CuCl)LaNb$_2$O$_7$, which are 
compatible with the NMR, NQR, and TEM results~\cite{Yoshida071} (Fig.~2).  In these models 
with orthorhombic distortion, 
displacement of Cl atoms generates different exchange couplings among the 
nearest neighbor Cu bonds, which were originally equivalent in the undistorted structure. 
The models shown in Fig.~2 where the strongest antiferromagnetic bonds are shown by 
the red (thick) lines then naturally lead to a dimer singlet groud state. 

To obtain further insight into the structure and the magnetism in (Cu$X$)LaNb$_2$O$_7$, we 
performed the NMR experiments in the collinear ordered phase of (CuBr)LaNb$_2$O$_7$. 
The internal field from ordered magnetic moments and the EFG at the nuclear sites provide 
useful information on the local structure.  We found a unique large internal field at both the 
Cu and the Br sites at 4.2~K, conforming a magnetic order.  However, if the Br nuclei 
coupled equally to the four neighboring Cu spins, the internal fields from these spins would have to be 
cancelled out to yield zero net field for the collinear AF structure.  Thus our results provide 
conclusive evidence for lower symmetry of the Br sites that allows the Br nuclei to couple 
dominantly to only two parallel Cu spins. 

\section{Experimental Results} 

The powder sample of (CuBr)LaNb$_2$O$_7$ was synthesized by 
the following ion-exchange reaction as described in Refs.~\citen{Kageyama051} and \citen{Oba061}, 
\begin{equation}
{\rm RbLaNb_2O_7} + {\rm CuBr_2} \rightarrow {\rm (CuBr)LaNb_2O_7} + {\rm RbBr} .  
\label{reaction}
\end{equation}
NMR spectra were obtained by recording the integrated intensity of the 
spin-echo signal at discrete frequencies. NMR spin-echo signal was observed 
at zero magnetic field and the temperature of 4.2 K in the frequency range 30-260~MHz. 

The obtained spectra are shown in Figs.~3(a) and 3(b) for the frequency 
ranges 80-260~MHz and 30-100~MHz, respectively.   As shown in Fig.~3(a), sharp 
six resonance lines are observed in the frequency range 100-260~MHz, which can be 
grouped into three pairs.  The peak frequencies are determined by fitting each line to a 
Lorentzian and the results are shown in Table I.  A natural interpretation for such a spectrum is the 
superposition of resonance from two kinds of spin 3/2 nuclei, both having the dominant magnetic 
Zeeman interaction split by the subdominant electric quadrupole interaction.  On the other hand, rather broad 
lines with only five visible peaks are observed in the frequency range 30-100~MHz, as shown 
in Fig.~3(b).  However, we have succeeded in decomposing the observed spectrum into 
three pairs of Lorentzian as shown by the red dashed lines.  Thus the spectrum below 
100~MHz can be interpreted again by two types of spin 3/2 nuclei with dominant magnetic 
Zeeman interaction and subdominant quadrupole interaction.  The peak frequencies in this 
frequency range are also listed in Table I.  There are indeed four spin 3/2 nuclear species 
contained in (CuBr)LaNb$_2$O$_7$, which are  $^{63}$Cu,  $^{65}$Cu, $^{79}$Br, 
and $^{81}$Br.  The values of their nuclear gyromagnetic ratio ($^{\alpha}\gamma$) 
and the isotopic ratios of the nuclear magnetic moments and the nuclear quadrupole moments 
($^{\alpha}Q$) are listed in Table II, where $\alpha$ stands for the mass number. 
\begin{figure*}[t]
\begin{center}
\includegraphics[width=0.9\linewidth]{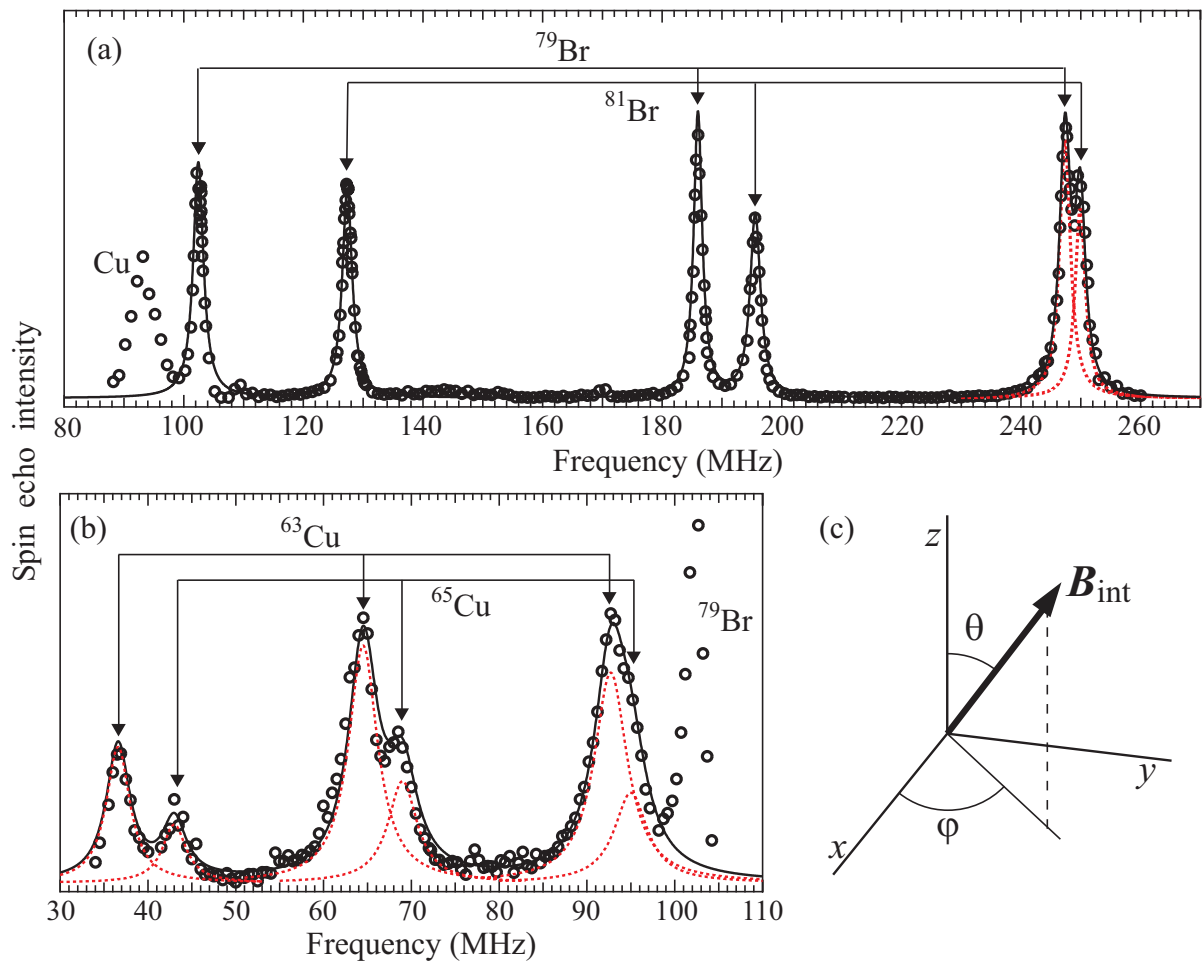}
\end{center}
\caption{(color online) The NMR spectra obtained at zero magnetic field and 4.2 K 
(open circles) for the frequency ranges (a) 80-260~MHz and (b) 30-110~MHz. 
The black solid lines show the fitting to the sum of six Lorentz functions and the red dashed 
lines represent the decomposition into each line.  (c) Definition of the polar and the azimuthal 
angles specifying the direction of the internal field with respect to the principal axes of the electric 
field gradient (EFG).}
\label{result}
\end{figure*} 

\section{Analysis I}

We first assign the resonance lines in Fig.~3(a)(b) to the four nuclear species. 
The NMR resonance frequencies for a spin  $I$ nucleus in zero magnetic field  
is generally determined by the following Hamiltonian~\cite{Slichter,Abragam}, 
\begin{eqnarray}
%\begin{split}
\mathcal{H} & = & \mathcal{H}_m + \mathcal{H}_q , \\
\mathcal{H}_m & = & -h ^{\alpha}\gamma \boldsymbol{I} \cdot \boldsymbol{B}_{\rm int}  \nonumber  \\ 
\mathcal{H}_q & = & \frac{h \nu_{Q}}{6} \left[ 3I_z^2 - I^2 + \frac{1}{2}\eta (I_+^2 + I_-^2) \right]   \nonumber
\label{hamilton}
%\end{split}
\end{eqnarray}
with 
\begin{equation}
\nu_{Q} = \frac{3e ^{\alpha} Q}{2I(2I-1)h}V_{zz},
\label{nuQ}
\end{equation} 
where $\mathcal{H}_m$ ($\mathcal{H}_q$) describes the magnetic 
Zeeman (electric quadrupole) interaction, $h$ is the Planck's constant, 
$\boldsymbol{B}_{\rm int}$ is the internal (hyperfine) magnetic field 
at the nuclear sites produced by the electronic magnetic moments, 
and $V_{zz} = \partial ^2V/\partial z^2$ is the $z$-component of the 
electric field gradient (EFG) tensor at the nucleus.  Here $x$, $y$ and 
$z$ denote the principal axes of the EFG tensor satisfying the relation 
$|V_{xx}| \leq  |V_{yy}| \leq  |V_{zz}|$ and the asymmetry parameter of 
the EFG is defined as $\eta  = (V_{xx} - V_{yy})/V_{zz}$ ($0 \leq \eta \leq 1$, 
note that $V_{xx} + V_{yy} + V_{zz} = 0$ ). 
If the quadrupole interaction is absent, the $2I+1$ nuclear spin levels 
are equally split by the Zeeman interaction, giving rise to a single resonance frequency 
\begin{equation}
\nu _{0} = \: ^{\alpha} \gamma B_{\rm int}. 
\end{equation} 
A finite value of the quadrupole 
coupling $\nu_{Q}$ causes the resonance to split into $2I$ lines. The 
frequencies of the split lines depend on the orientation of the internal 
magnetic field relative to the principal axes of the EFG.  We define the 
polar and the azimuthal angles $\theta$ and $\phi$ specifying the direction 
of $\boldsymbol{B}_{\rm int}$ with respect to the principal axes 
of the EFG as shown in Fig.~3(c). 
\begin{table}[t]
\caption{Peak frequencies of the NMR spectra in Fig.~3.}
\label{t1}
\begin{tabular}{cccc}
\hline
  & $\nu _{\rm sl}$ (MHz) & $\nu _{\rm c}$ (MHz) & $\nu _{\rm sh}$ (MHz) \\
\cline{2-4}
$^{63}$Cu & 36.6 $\pm $ 0.3 & 64.5 $\pm $ 0.3 & 92.7 $\pm $ 0.5 \\
$^{65}$Cu & 43.0 $\pm $ 0.5 & 69.0 $\pm $ 0.5 & 95.0 $\pm $ 1.0 \\
$^{79}$Br & 102.44 $\pm $ 0.1 & 185.98 $\pm $ 0.05 & 247.36 $\pm $ 0.2 \\
$^{81}$Br & 127.34 $\pm $ 0.1 & 195.56 $\pm $ 0.1 & 249.9 $\pm $ 0.2 \\
\hline
\end{tabular}
\end{table} 
\begin{table}[t]
\caption{Values and isotopic ratios of nuclear moments \cite{Bieron}}
\label{t2}
\begin{tabular}{ccccc}  \hline
Cu & $^{63}\gamma$ (MHz/T) & $^{65}\gamma$ (MHz)/T & $^{65}\gamma /^{63}\gamma $ & $^{65}Q/^{63}Q$ \\  
%\cline{2-5}
     & 11.2893 & 12.0932 & 1.0712 & 0.9252 \\
\hline
Br & $^{79}\gamma$ (MHz/T) & $^{81}\gamma$ (MHz)/T & $^{81}\gamma /^{79}\gamma $ & $^{81}Q/^{79}Q$ \\  
%\cline{2-5}
     & 10.6671 & 11.4984 & 1.0779 & 0.8354 \\
\hline
\end{tabular}
\end{table}

When the magnetic Zeeman interaction is dominant over the quadrupole 
interaction, $^{\alpha}\gamma B_{\rm int} \gg \nu_{Q}$, which indeed turns 
out to be case for the NMR spectra in Fig.~3, the resonance frequencies 
can be calculated by the perturbation theory up to second order with 
reasonably good accuracy. For the case of $I$ = 3/2, there are three 
resonance lines; the central line, the high frequency satellite and the low 
frequency satellite lines, whose frequencies are denoted by 
$\nu_{\rm c}$, $\nu_{\rm hs}$, and $\nu_{\rm ls}$. The perturbation theory gives the 
following expression,
\begin{subequations}
\begin{align}
&\nu_{\rm c} = \nu_{0} + \delta \nu_{\rm c}^{(2)}  \\
&\nu_{\rm hs} = \nu_{0} + \delta \nu ^{(1)} + \delta \nu_{\rm s}^{(2)}  \\
&\nu_{\rm ls} = \nu_{0} - \delta \nu ^{(1)} + \delta \nu_{\rm s}^{(2)}.
\end{align}
\end{subequations}
The first order shift $\delta \nu ^{(1)}$ and the second order shifts $\delta \nu _{c}^{(2)}$ 
and $\delta \nu _{s}^{(2)}$ are expressed as \cite{Stauss}  
\begin{subequations}
\begin{align}
\delta \nu ^{(1)} =& (\nu _Q/2) \mid 3\cos^2\theta -1 + \eta \cos2\phi \sin^2\theta \mid \\ 
\delta \nu_{\rm c}^{(2)} =& (\nu _Q^2/12\nu _0)\{ - (9/4) \sin^2\theta (9\cos^2\theta  - 1) \notag \\ 
& + (3/2)\eta \cos2\phi \sin^2\theta (9\cos^2\theta + 1) \notag \\ 
& + \eta ^2(- 2 + 3\cos^2\theta + (9/4)\cos^22\phi \sin^4\theta)\} \\ 
\delta \nu_{\rm s}^{(2)} =& (\nu _Q^2/12\nu _0)\sin^2\theta \{ 18\cos^2\theta 
- 12\eta \cos2\phi \cos^2\theta \notag \\ 
&+ 2\eta ^2(1 - \cos^22\phi \sin^2\theta)\}.
\end{align}
\end{subequations}
In general, the $n$-th order shifts $\delta\nu_{\rm c, hs, ls}^{(n)}$ for the three resonance lines 
satisfy the relations, $\delta\nu_{\rm c}^{(n)}$ = 0 and $\delta\nu_{\rm hs}^{(n)} = - \delta\nu_{\rm ls}^{(n)}$ for
odd $n$ and $\delta\nu_{\rm hs}^{(n)} = \delta\nu_{\rm ls}^{(n)}$ for even $n$. 

From Eqs. (3), (5b), (5c) and (6a), one finds that the separation between 
the two satellite lines is proportional to the nuclear quadrupole moment, 
\begin{equation}
\nu_{\rm hs} - \nu_{\rm ls} = 2\delta \nu ^{(1)} \propto \: ^{\alpha} Q.
\end{equation} 
This means that the ratio of this separation for the two isotopes on the same chemical 
and magnetic environment is equal to the ratio of their nuclear quadrupole moments. 
The NMR spectra in the two panels of Fig.~3 both show two sets of quadrupole split 
three lines.  The ratio of the difference between the highest and the lowest frequencies 
obtained from the data in Table I is 0.8457 for the spectrum in (a) and 0.9269 
for the spectrum in (b). These values are indeed very close to the isotopic ratio 
of the nuclear quadrupole moments for Br (0.8354) and Cu (0.9251) shown in the Table II, 
allowing us to assign the resonance lines as indicated in Fig.~3 and Table I. 
The first order quadrupole shifts are obtained as 
\begin{subequations}
\begin{align}
\delta \nu ^{(1)}& = 28.1 \  (26.0) \ {\rm MHz \ for \ ^{63}Cu \ (^{65}Cu)}  \\
\delta \nu ^{(1)}& = 72.46 \ (61.28) \ {\rm MHz \ for \ ^{79}Br \ (^{81}Br)}. 
\end{align}
\end{subequations} 
These values are much smaller than the Zeeman frequencies, which is approximately 
equal to the center line frequency shown in Table I. Therefore, our use of the perturbation 
theory is justified. 

If we neglect the second order quadrupole shift, the magnitude of the internal field 
$B_{\rm int}$ can be crudely estimated from the center line frequency as 
$B_{\rm int} \approx 5.7$~T for the Cu sites and $B_{\rm int} \approx 17$~T 
for the Br sites.  More accurate determination is described 
in the next section. The existence of such large internal fields at zero magnetic field 
is the direct evidence for a magnetically ordered state with spontaneous 
magnetic moments, consistent with the neutron diffraction experiments~\cite{Oba061}. 

\section{Analysis II} 

In this section, we try to determine the internal field, the EFG, and their relative orientation 
from the NMR frequencies for both the Br and the Cu sites.  The nuclear Hamiltonian Eq.~(2) 
is specified by five parameters, $B_{\rm int}$, $\nu_{Q}$, $\eta$, $\theta$, and $\phi$. 
Their values are independent of isotopes except for $\nu_{Q}$, which is proportional to the nuclear 
quadrupole moment (Eq. 3). Although the second order perturbation expression, Eqs.~(5a)-(5c) 
with Eqs.~(6a)-(6c),  gives reasonably accurate resonance frequencies, it contains only four 
independent parameters, $\nu_{0}$, $\nu^{(1)}$, $\nu_{\rm c}^{(2)}$, and $\nu_{\rm s}^{(2)}$. 
Therefore, we must examine the higher order terms to determine all the parameters of the Hamiltonian. In what 
follows, we first use the perturbation expression to find an appropriate range of the parameter 
values. We then diagonalize the Hamiltoninan Eq.~(2) numerically to calculate the resonance frequencies 
exactly and seek the best parameter values which minimizes $\chi^{2}$ defined as 
$\chi ^2 = \sum_{\alpha,k} \left[ \left( ^{\alpha}\nu^{obs}_{k} - ^{\alpha}\nu^{cal}_{k} \right)^2/^{\alpha}\sigma_{k} ^2 \right]$, 
where $ ^{\alpha}\nu^{obs}_{k}$ ($^{\alpha}\nu^{cal}_{k}$) is the observed (calculated) resonance 
frequency of the $k$-th line ($k$ = c, hs, or ls) for the isotope $\alpha$ and $^{\alpha}\sigma_{k}$
is the experimental uncertainty listed in Table I. 

For the case of Br spectrum showing sharp lines, 
the frequencies can be determined quite accurately. We found that nearly a unique set of parameters 
can be obtained already by the perturbation analysis. For the case of Cu showing rather broad spectrum, 
however, we were not able to determine the parameters completely. 

\subsection{Br sites} 

We first determine each term in the perturbation expression from the observed resonance 
frequencies for the two Br isotopes. From Eqs.~(4) and (6a)-(6c), the values of each term in the 
perturbation expression for the two isotopes can be related as
\begin{subequations}
\begin{align} 
^{81}\nu_{0} =&~\left( ^{81}\gamma /^{79}\gamma \right) \,^{79}\nu _{0}=1.0779 \times \,^{79}\nu _{0}  \\
^{81}\delta\nu^{(1)} =&~\left( ^{81}Q/^{79}Q \right) \,^{79}\delta\nu^{(1)}=0.8354 \times \,^{79}\delta\nu^{(1)}  \\
^{81}\delta\nu_{\rm c,s}^{(2)} =&~\frac{\left( ^{81}Q/^{79}Q \right)^2}{\left( ^{81}\gamma /^{79}\gamma \right)} \,^{79}\delta\nu_{\rm c,s}^{(2)}=0.6474 \times \,^{79}\delta\nu_{\rm c,s}^{(2)}.
\end{align}
\end{subequations}
From the observed values of $^{79}\nu_{\rm c}$ and $^{81}\nu_{\rm c}$ (Table I), we obtain 
 $^{79}\nu_{0}$ and $^{79}\delta\nu_{\rm c}^{(2)}$, using Eqs.~(5a), (9a) and (9c), 
\begin{equation}
^{79}\nu_{0} = 174.57 \pm 0.14 \ {\rm MHz},  \ \ ^{79}\delta\nu_{\rm c}^{(2)} =  11.41 \pm 0.17 \ {\rm MHz}. 
\end{equation} 
Then from the observed values of $^{79}\nu_{\rm ls}$ and $^{79}\nu_{\rm hs}$ 
and Eq.~(5b) and (5c), we obtain   
\begin{equation}
^{79}\delta\nu^{(1)} = 72.46 \pm 0.14 \ {\rm MHz},  \ \ ^{79}\delta\nu_{\rm s}^{(2)} = 0.33 \pm 0.14 \ {\rm MHz}. 
\end{equation} 
%One may also determine these parameters from the data for $^{81}$Br, using the relation Eqs.~(9b)
%and (9c) to obtain 
%\begin{equation}
%^{79}\delta\nu^{(1)} = 73.35 \ {\rm MHz}, \ \ ^{79}\delta\nu_{\rm s}^{(2)} = 0.69 \ {\rm MHz}. 
%\end{equation} 
%The small differences between Eqs.~(11) and (12) should be ascribed to the third and the higher order 
%terms in perturbation.  

%\begin{figure}[tb]
%\begin{center}
%\includegraphics[width=0.8\linewidth]{fig4}
%\end{center}
%\caption{$\beta $-dependence of $c'$ for various $\eta$ and $\alpha $.}
%\label{ana1}
%\end{figure}  
We now use the above results to find the possible range of parameter values in the nuclear 
Hamiltonian.  Note that the second order shift for the satellite lines $^{79}\delta\nu_{\rm s}^{(2)}$ (Eq.~11) is 
very small compared with that for the central line $^{79}\delta\nu_{\rm c}^{(2)}$ (Eq.~10). This puts strong constraints 
as we see below. First, by noting that the upper limit of the right-hand-side of Eq.~(6a) 
is set by $\nu_{Q}$, we obtain $^{79}\delta\nu^{(1)} \leq \,^{79}\nu _{Q}$.  The experimental value 
of $^{79}\delta\nu^{(1)}$ in Eq.~(11) then sets a lower limit for $^{79}\nu _{Q}$, $^{79}\nu _{Q} \geq 72.46$~MHz. 
Next, putting this lower limit and the experimental value of $^{79}\nu_{0}$ in Eq.~(10) into Eq.~(6c), we obtain 
\begin{equation}
^{79}\delta \nu_{\rm s}^{(2)} \geq 2.506 (1 - \mu) \{ ( 18 - 12 \eta \alpha + 2 \eta^{2} \alpha^{2} ) \mu + 2 \eta^{2}( 1 - \alpha^{2}) \},
\end{equation} 
where we defined $\mu = \cos^2\theta$ and $\alpha = \cos2\phi$. Note that the minimum of the right-hand-side of Eq.~(12) 
is located at $\eta$ = 1 and $\alpha$ = $1$ within the defined range $0 \leq \eta \leq 1$, $-1 \leq \alpha \leq 1$ for any value of $\mu$. Therefore, 
\begin{equation}
^{79}\delta \nu_{\rm s}^{(2)} \geq 20.05 \mu (1-\mu). 
\end{equation}
Then the small experimental value of $^{79}\delta \nu_{\rm s}^{(2)}$ = 0.33~MHz restricts 
the possible range of $\mu$ either near zero ($\mu \leq 0.017$, $\theta \geq 82^{\circ}$) 
or near one ($\mu \geq 0.983$, $\theta \leq 7.4^{\circ}$). If $\mu \simeq 1$ ($\theta \simeq 0$), 
$^{79}\nu _{Q} \simeq \,^{79}\delta\nu^{(1)} = 72.46$~MHz from Eqs.~(6a) and (11). 
Then from Eq.~(6b), $^{79}\delta \nu _{\rm c}^{(2)} \simeq 2.506 \eta^2$~(MHz). However, this is 
clearly incompatible with the experimental value of 11.41~MHz (Eq.~10). Therefore, possible solutions 
must satisfy $\mu \simeq 0$ ($\theta \simeq 90^{\circ}$).  

The possible ranges of parameter values are determined as follows. We first fix the value of $\mu$. 
Then $^{79}\nu _{Q}$ is expressed by $\eta$ and $\alpha$ using Eq.~(6a) and the experimental value of 
$^{79}\delta\nu^{(1)}$.  By putting this expression into Eqs.~(6b) and (6c), the values of 
$\eta$ and $\alpha$ are determined by solving these equations using the experimental values of 
$^{79}\delta\nu_{\rm c}^{(2)}$ and $^{79}\delta\nu_{\rm s}^{(2)}$.  We found that solutions exist only 
for $0 \leq \mu \leq 0.0024$ ($90^{\circ} \leq \theta \leq 87.2^{\circ}$).  For this range of  $\theta$, 
other parameters take the values, $-0.85 \leq \alpha \leq -1$ ($74^{\circ}\leq \phi \leq 90^{\circ}$), $0.31 \geq \eta \geq 0.28$, 
and $114.7 \leq \,^{79}\nu _{Q} \leq 115.1$~MHz. 
%If we use the experimental values in Eqs.~(10) and (12) instead, solutions can be found between
%$\theta$=90$^{\circ}$, $\phi$=21$^{\circ}$,  $\eta$=0.35, $^{79}\nu _{Q}$=116.3~MHz
%and 
%$\theta$=86$^{\circ}$, $\phi$=0$^{\circ}$,  $\eta$=0.27, $^{79}\nu _{Q}$=117.2~MHz. 
The magnitude of the internal field is obtained as $B_{\rm int}= \,^{79}\nu_{0}/^{79}\gamma = 16.37$~T. 
Thus the second order perturbation analysis significantly narrows down the possible range of the parameters. 

\begin{table}[t]
\caption{Calculated Br NMR frequencies. The numbers in parentheses indicate the difference from the experimental values
shown in Table I.}
\label{t3}
\begin{tabular}{cccc}
\hline
  & $\nu _{\rm sl}$ (MHz) & $\nu _{\rm c}$ (MHz) & $\nu _{\rm sh}$ (MHz) \\
\cline{2-4}
$^{79}$Br & 102.53  (0.09) & 185.97  (-0.01) & 247.38  (0.02) \\
$^{81}$Br & 127.31   (-0.03) & 195.64  (0.08)  & 249.85  (-0.05) \\
\hline
\end{tabular}
\end{table} 
We have further refined the parameter values obtained above by numerical diagonalization of the 
nuclear Hamiltonian to calculate $\chi ^2$ and by seeking the minimum of $\chi ^2$. 
We have changed the value of each parameter successively within an appropriate range to find the global 
minimum of $\chi ^2$ by iteration. 
%$16 \leq B_{\rm int} \leq 17$~T, $100 \leq ^{79}\nu_Q \leq 130$~MHz,
%$0^{\circ} \leq \theta \leq 90^{\circ}$, $0^{\circ} \leq \phi \leq 90^{\circ}$, $0 \leq \eta \leq 1$.  
Then the final adjustment of the parameter values was done by using parabolic expansion of $\chi ^2$ with respect 
to the parameters to calculate the precise location of the minimum and to estimate uncertainties of the parameter values. 
The final results are the following: 
\begin{eqnarray}
B_{\rm int} & = & 16.40 \pm 0.01 \ \ ({\rm T})  \nonumber \\
^{79}\nu _Q & =  & 112.5 \pm 0.4 \ \ ({\rm MHz}) \nonumber \\
\eta & = & 0.330 \pm 0.005   \nonumber \\
\theta & = & 89.5^{\circ} \pm 1.0^{\circ}   \nonumber \\
\phi  & = & 87.8^{\circ} \pm 4.9^{\circ} ,
\end{eqnarray}  
with $\chi^2$ = 1.8. The resonance frequencies calculated from these parameter values 
are compared with the experimental data in Table III.  The finite value of $\eta$ indicates 
lack of $C_{4}$-symmetry at the Br sites, similarly to the Cl sites in (CuCl)LaNb$_2$O$_7$ 
where $\eta$ = 0.56~\cite{Yoshida071}, again ruling out the tetragonal $P4/mmm$ space group. 
 
\subsection{Cu sites} 

Applying the same perturbation analysis to the Cu data, we obtain the following values of the 
Zeeman frequency, the first and the second order quadrupolar shifts for the $^{63}$Cu nuclei, 
$^{63}\nu_{0}$ = 64.2 $\pm$ 2.0~MHz ($B_{\rm int}$ = 5.7~T), $^{63}\delta\nu^{(1)}$ = 28.05 $\pm$ 0.29~MHz, 
$^{63}\delta\nu_{\rm c}^{(2)}$ = 0.3 $ \pm$ 2.1~MHz, $^{63}\delta\nu_{\rm s}^{(2)}$ = 0.5 $\pm$ 2.1~MHz. 
The rather large uncertainties of these values, however, prevent us from narrowing the 
range of parameter values efficiently. 

\begin{figure}[tb]
\begin{center}
\includegraphics[width=0.9\linewidth]{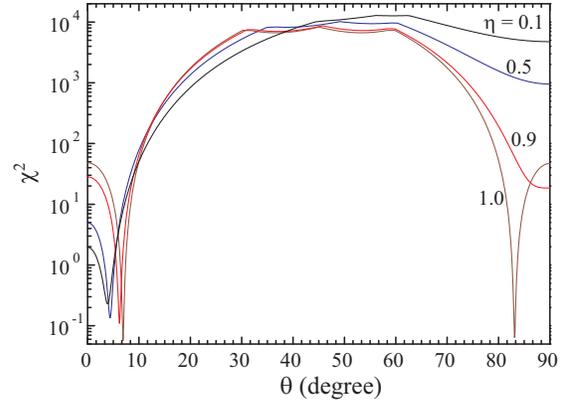}
\end{center}
\caption{(color online). The $\theta$-dependence of $\chi ^2$ of the Cu NMR spectrum for various values of $\eta$. 
Values of other parameters are chosen to minimize $\chi ^2$: 
$B_{\rm int}$ = 5.716~T, $^{63}\nu_Q$ = 28.31~MHz, $\phi$ = 1.3$^{\circ}$ for $\eta$ = 0.1, 
$B_{\rm int}$ = 5.705~T, $^{63}\nu_Q$ = 28.45~MHz, $\phi$ = 89.8$^{\circ}$ for $\eta$ = 0.5,
$B_{\rm int}$ = 5.680~T, $^{63}\nu_Q$ = 29.03~MHz, $\phi$ = 90$^{\circ}$ for $\eta$ = 0.9
and $B_{\rm int}$ = 5.670~T, $^{63}\nu_Q$ = 29.38~MHz, $\phi$ = 88.7$^{\circ}$ for $\eta$ = 1.0.
Note that $\chi ^2$ always takes one deep minimum at a small value of $\theta$ ($\theta \leq 6^{\circ})$
except when $\eta$ is very close to one.}
\label{ana1}
\end{figure}  
In fact, the experimental Cu resonance frequencies can be reproduced quite well by numerical 
diagonalization of the nuclear Hamiltonian for any value of $\eta$ between 0 and 1 as shown in 
Fig.~4, where $\chi ^2$ is plotted against $\theta$ for several values of $\eta$ with other parameter 
values fixed to minimize $\chi ^2$.  The $\chi ^2$ always shows a single deep minimum well 
below one at a small value of $\theta$, $\theta \leq 7^{\circ}$, except when $\eta$ is very 
close to one. For $\eta$ = 1, there are two minima in $\chi^2$ at $\theta$ = 6$^{\circ}$ and at 
$\theta$ = 83$^{\circ}$. The optimized values of $B_{\rm int}$ and $^{63}\nu_Q$ minimizing 
$\chi ^2$ depend only slightly on $\eta$: $B_{\rm int}$ = 5.72~T and $^{63}\nu_Q$ = 28.3~MHz 
for $\eta$ = 0 change to $B_{\rm int}$ = 5.67~T and $^{63}\nu_Q$ = 29.4~MHz for $\eta$ = 1. 
The optimized value of $\phi$ stays at 90$^{\circ}$ when $\eta \geq 0.3$. When 
$\eta$ is small ($\eta \simeq 0.1$), $\phi$ cannot be determined since the Hamiltonian 
itself depends only weakly on $\phi$.      

Let us compare these results with those obtained for the non-magnetic (CuCl)LaNb$_2$O$_7$ 
by Cu NMR experiments at various magnetic fields~\cite{Yoshida071}.  By assuming that the 
$c$-axis is one of the principal axes of the EFG, it was found that the $z$-axis with the 
largest principal value of EFG in (CuCl)LaNb$_2$O$_7$ lie in the $ab$-plane, 
$^{63}\nu_Q$ = 29.64~MHz and $\eta$ = 0.098~\cite{Yoshida071}. We notice that 
the values of $^{63}\nu_Q$ are almost identical for the two materials, indicating the 
similar local structure.  We also expect that they have similar values of $\eta$. Therefore, 
it is very unlikely that $\eta$ in (CuBr)LaNb$_2$O$_7$ is close to one. Then we conclude 
from the above results that $\theta$ is very small ($\theta \leq 6^{\circ}$) in (CuBr)LaNb$_2$O$_7$. 
The analysis of the Cu spectrum in (CuBr)LaNb$_2$O$_7$demonstrates that the Cu-NMR spectrum 
can be well reproduced by the EFG parameters which are nearly equal to those for (CuCl)LaNb$_2$O$_7$ 
and the internal field of about 5.7~T must be directed nearly parallel to the $z$-axis of the EFG tensor, 
which corresponds to the largest principal value of EFG. 

\section{Discussion} 

Summarizing the analysis in the previous sections, we found the following: (1) There is an internal 
field of 5.7~T at the Cu sites directed along the $z$-axis of the EFG.  (2) There is an internal field 
of 16.4~T at the Br sites directed perpendicular to the $z$-axis of the EFG. 

Let us first discuss the results on the Cu sites. Since the value of $\nu_Q$ at the Cu site is nearly the 
same for the Br and Cl compounds, we expect the two compound to have the same $z$-axis of the 
EFG, which was found to be in the $ab$-plane in (CuCl)LaNb$_2$O$_7$~\cite{Yoshida071}. 
The internal field directed along the $z$-axis in (CuBr)LaNb$_2$O$_7$ is then consistent with 
the neutron results reporting the ordered moment parallel to the $b$-axis~\cite{Oba061}. 
The internal field at the Cu nuclei is mainly 
due to the magnetic moment on the same Cu ion being observed (the on-site hyperfine field), 
which is anisotropic reflecting the anisotropic shape of the Cu-3d orbital. 
It is concluded from the anisotropy of the Cu NMR shift in (CuCl)LaNb$_2$O$_7$ that the Cu 
spin is mainly on the $d(3y^2-r^2)$ orbital, where $y$ is parallel to the crystalline $c$-axis~\cite{Yoshida071}. 
The internal field of 5.7~T then corresponds to the magnetic moment of 0.5$\pm$0.1~$\mu_{B}$ 
in the $ab$-plane from the hyperfine coupling constant $A^{\rm hf}_{\perp} = 
-11 \pm 2$~T/$\mu_{B}$~\cite{Yoshida071}.  This value is consistent with 
the ordered moment of 0.6~$\mu_{B}$ observed by neutron diffraction~\cite{Oba061}. 

\begin{figure}[tb]
\begin{center}
\includegraphics[width=0.8\linewidth]{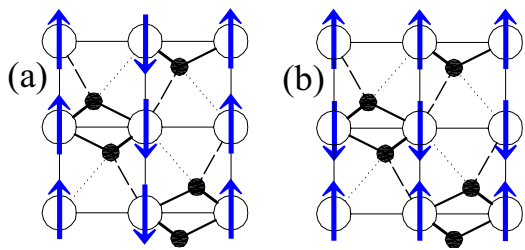}
\end{center}
\caption{(color online). Possible spin configurations of the collinear order on the distorted lattice. Only the pattern (b) is 
compatible with the large internal field observed at the Br sites.}
\label{ana1}
\end{figure}  
We next discuss the results on the Br sites. In contrast to the Cu sites, the internal field at the Br sites 
is due to the transferred hyperfine interaction, which is the coupling between the magnetic moments 
and the ligand nuclei through the covalency effects. If the structure were the tetragonal $P4/mmm$~\cite{Kodenkandath991}, 
the Br nuclei would couple equally to the four nearest neighbor Cu moments.  Then the total internal field 
would have to be cancelled out to zero for the collinear AF order with the wave vector $(\pi, 0, \pi)$, which was 
observed by neutron diffraction experiments~\cite{Oba061}. The finite internal field at the Br sites, 
therefore, gives direct evidence for structural distortion breaking the four-fold symmetry at the Br sites. 
The finite value of $\eta$, the asymmetry parameter of EFG, at the Br sites is another independent 
piece of evidence for such structural distortion. Yet, the sharp six peaks in the Br-NMR spectrum 
ensure a unique crystallographic site without appreciable disorder. 

These features of the local symmetry at the Br sites are exactly what were found for the Cl sites in 
(CuCl)LaNb$_2$O$_7$~\cite{Yoshida071}, which have led Yoshida {\it et al.} to propose possible 
structural models with orthorhombic distortion shown in Fig.~2. 
We now examine if the same models provide consistent 
account for the results on (CuBr)LaNb$_2$O$_7$ as well. In (CuCl)LaNb$_2$O$_7$, the $z$-axis 
of the EFG at the Cl sites is parallel to the $c$-axis. This is likely to hold also for the Br sites in 
(CuBr)LaNb$_2$O$_7$, since the internal field at the Br sites should be parallel to the Cu moments, 
which lie in the $ab$-plane, and $\theta$ = 90$^{\circ}$. This prefers the model (a) 
rather than (b) and (c) shown in Fig.~2, since the Cl (or Br) sites are then on the mirror plane, therefore, the 
$c$-axis can be a principal axis. Yoshida {\it et al.} have argued that the displacement of 
Cl atoms makes the exchange coupling shown by the red lines in Fig.~2 much stronger 
than the other bonds, leading to local dimer singlet formation in (CuCl)LaNb$_2$O$_7$. 

If this bond remains strong and 
antiferromagnetic in (CuBr)LaNb$_2$O$_7$, the collinear AF order shown in Fig.~5(a) should 
appear. In this structure, however, Br nuclei couple dominantly to two antiparallel moments, therefore,
the internal field can never be as large as 16.4~T. In fact, the observed internal field is nearly 
equal to those reported for (CH$_3$NH$_3$)$_2$Cu(Cl$_{1-x}$Br$_{x}$)$_4$ 
(15.8 and 15.2~T)~\cite{Kubo791}, which is a typical example of strong transferred hyperfine 
interaction via the $\sigma$-bond between Cu-$d(x^{2}-y^{2})$ orbital and the Br-$p_{\sigma}$ orbital.   
The transferred hyperfine interaction should be weaker for a Br-Cu pair in (CuBr)LaNb$_2$O$_7$,  
where the spin density is mainly on the $d(3y^2-r^2)$ orbital. Thus the large 
internal field of 16.4~T is possible only if the hyperfine fields from two Cu spins add.  In other words,
the Br nuclei must couple dominantly to two parallel moments as shown in Fig.~5(b).  

The same hybridization process involved in the transferred hyperfine interaction also 
causes the superexchange interactions. The strong asymmetry in the transferred 
hyperfine interaction, then induces inequivalent exchange interactions. The models 
shown in Fig. 2(a), for example, include five distinct exchange interactions.      
It is known that the superexchange interaction through the path Cu-X-Cu is very sensitive
to small structural variation and often changes sign~\cite{Inagaki051}.  Therefore,
it is conceivable that dominant exchange coupling changes from antiferromagnetic in 
(CuCl)LaNb$_2$O$_7$ to ferromagnetic in (CuBr)LaNb$_2$O$_7$ by small 
difference in the ionic position, even if the both compounds show the same type of 
structural distortion.   

From the susceptibility and the magnetization data, however, it is unlikely that single 
ferromagnetic interaction dominate over other interactions in (CuBr)LaNb$_2$O$_7$.  
The Weiss temperature $\theta_{W}$, defined by the Curie-Weiss 
law of the susceptibility $\chi = C/(T+\theta_{W})$, changes from 9.6 K in 
(CuCl)LaNb$_2$O$_7$~\cite{Kageyama051} to 5.1 K in 
(CuBr)LaNb$_2$O$_7$~\cite{Oba071}. The positive value of $\theta_{W}$ indicates that averaged interaction 
is still antiferromagnetic in (CuBr)LaNb$_2$O$_7$, even though it is less so compared with 
(CuCl)LaNb$_2$O$_7$. The saturation field $B_{c}$ of the magnetization curve gives a 
measure of stability of the antiferromagnetic state. The icrease of $B_{c}$ from 30 T in 
(CuCl)LaNb$_2$O$_7$~\cite{Kageyama052} to 70 T in (CuBr)LaNb$_2$O$_7$~\cite{Oba071} 
points to higher stability of the antiferromagnetic state in (CuBr)LaNb$_2$O$_7$, 
presumably due to overall increase of exchange interaction in Br compounds.  
It appears that the different ground states in the Cl and Br compounds is the results 
of subtle balance among several inequivalent frustrating interactions. Precise theoretical 
studies are desired to relate our structural model to the magnetic ground state.   

\section{Summary} 
We have presented the Cu and Br NMR spectra in the collinear antiferromagnetic 
state in (CuBr)LaNb$_2$O$_7$ at zero magnetic field and $T$ = 4.2~K and determined the values of 
the internal field, EFG tensor and their relative orientation. The results at the Cu sites 
confirmed an AF order of about 0.5$\pm$0.1~$\mu_{B}$/Cu in the $ab$-plane. 
In order for the large internal field at the Br sites to be compatible with the collinear AF order 
observed by neutrons, the Br sites must be displaced from the center of the Cu square lattice 
to couple strongly to only two parallel Cu spins.  Our results in (CuBr)LaNb$_2$O$_7$ 
can be explained consistently by the structural model proposed for (CuCl)LaNb$_2$O$_7$, 
provided that the pairs of Cu spins, which form singlet dimes in the Cl compound, are aligned parallel 
in the Br compounds due to modified exchange interactions brought by small difference in the ionic 
displacement. 

\section*{Acknowledgment} 
This work was supported by Grant-in-Aid for Scientific Research (Nos. 18740202 and 17684018) 
and Grant-in-Aids on Priority Areas ``Invention of Anomalous Quantum Materials'' (Nos. 16076204 
and 16076210) and ``Novel State of Matter Induced by Frustration'' (No. 19052004) from the 
MEXT Japan.

\end{document}